\documentclass[aps,prl,psfig,twocolumn,floats,showpacs]{revtex4}
\usepackage{psfig}

\begin{document} 

\title{Ultrafast light-induced magnetization dynamics in ferromagnetic
semiconductors} 

\author{J. Chovan, E. G. Kavousanaki, and 
I. E.  Perakis}

\affiliation{Institute of Electronic Structure \& Laser, Foundation
for Research and Technology-Hellas
and Department of Physics, University of Crete, Heraklion, Greece}

\date{\today}

\begin{abstract}
We develop a theory of the magnetization
dynamics triggered by ultrafast optical excitation of 
ferromagnetic semiconductors.
We describe the effects of the strong carrier spin relaxation 
on the nonlinear optical response 
by using the Lindblad semigroup 
method. 
We demonstrate magnetization control 
during femtosecond timescales via the 
interplay between circularly polarized optical excitation, 
hole--spin damping, polarization dephasing, and the Mn--hole spin
interactions. Our results show a light--induced
magnetization precession and relaxation for the duration of the optical pulse.

\end{abstract}


\maketitle 


With the  discovery of III-Mn-V 
ferromagnetic semiconductors \cite{ohno}, 
one can envision 
a new class of multifunctional devices that combine
information processing and  storage
on a single chip with low power consumption. 
One of the challenges facing such future
devices concerns the speed of the basic processing unit, 
governed by  the 
 dynamics of the collective spin
and its ability 
to respond quickly to time-dependent external fields.
Ultrafast pump--probe magneto--optical spectroscopy 
has been used to study the magnetization dynamics triggered by 
femtosecond optical pulses 
in both metals \cite{bigot,koop}
and  ferromagnetic semiconductors 
\cite{kimel-04,wang-05,kojima-03,oiwa}. 
Transient magnetic effects 
have been observed even in the  initial nonthermal
temporal regime, where 
the concept of carrier temperature is not meaningful
\cite{bigot,gomez}.

In III-Mn-V semiconductors, the  ferromagnetic order 
is induced by the 
interaction between the itinerant  valence band 
hole spins and the 
localized Mn spins \cite{mean,review}. 
The magnetization is therefore 
sensitive to the itinerant carrier properties, which 
in III-V semiconductors can be well controlled 
with ultrashort optical pulses  \cite{sbe}. 
The combined magnetic and optical properties 
of this system 
open new possibilities for manipulating 
the collective spin without using magnetic fields. 
This can be achieved, e.g., by
exciting spin--polarized   carriers
with circularly--polarized femtosecond  optical 
pulses. The effects of such photoexcited 
spin on the ferromagnetic state 
can be studied by monitoring    the change in the  macroscopic magnetization
as  function of time using ultrafast  spectroscopy 
\cite{bigot,koop,kimel-04,wang-05,kojima-03,oiwa}. 
Light--induced magnetization and 
very fast 
 Mn spin  rotation 
towards the direction of propagation of 
the circularly polarized  optical pulse, 
perpendicular to the ground state magnetization,  
were reported in
 Ga(Mn)As ferromagnetic epilayers 
 \cite{oiwa}.
Kimel et.al. \cite{kimel-04} 
observed a 
photoinduced magnetization 
attributed 
to the 
photoexcited 
conduction electron spin.
Kojima et.al. \cite{kojima-03}
interpeted their data in terms of  a rapid (ps) rise in carrier temperature, 
followed by a slow rise in the spin temperature 
and a magnetization reduction 
over 100s of ps. 
For very intense photoexcitation of In(Mn)As,
Wang et.al. \cite{wang-05} observed 
a fast (100s of fs) and a slow (100s of ps)
 demagnetization regime.
Other possibilities for light--induced 
magnetic effects in non-magnetic semiconductors 
were recently explored theoretically \cite{kondo,pier-04}.  
In doped semiconductors, 
a light--induced Kondo effect 
was predicted \cite{kondo}, 
while  a light-induced paramagnetic-to-ferromagnetic
phase transition was predicted in undoped semiconductors \cite{pier-04}.

In this letter, 
we develop 
a theory 
of the ultrafast nonlinear
optical dynamics and light-induced hole and Mn spin relaxation 
in ferromagnetic semiconductors. 
We treat 
the magnetic exchange interactions within the 
time--dependent Hartree--Fock approximation,  
and treat the hole spin damping and resulting nonlinear polarization dephasing 
within the Lindblad 
semigroup method \cite{lindblad}.
We predict 
an ultrafast  light--induced precession  and relaxation 
of the collective magnetization towards a direction determined 
by the nonlinear optical polarizations.
The above spin dynamics 
results from the interplay between the 
nonlinear optical excitation,  
the interaction between the 
photoexcited hole and Mn spins, 
and the strong hole--spin damping and optical polarization dephasing, 
The predicted  dynamics  is observable with 
ultrafast magneto--optical pump--probe spectroscopy.

We adopt
the simple \cite{mean,review} two--band sp--d  model Hamiltonian 
$ H(t)=K_e + K_h + H_{\rm{exch}} +H_{L}(t)$, 
where 
$K_h$ ($K_e$) is  the kinetic energy of the valence band heavy holes 
(conduction band electrons) 
created by  $\hat{h}_{{\bf k} \sigma}^{\dag}$
($\hat{e}_{{\bf k}\sigma}^{\dag}$) 
with total angular momentum $J_z=\pm 3/2$ 
(spin  $\sigma=\pm 1/2$) 
and 
dispersion  $\varepsilon^{v}_{{\bf k}} =  k^2/2 m_h$
($\varepsilon^{c}_{{\bf k}}=  k^2 / 2 m_e$) 
(we set $\hbar$=1).  
Unlike in II-VI semiconductors, 
the Mn impurities provide 
both  a hole Fermi sea  and 
randomly distributed  $S=5/2$  
localized  spins, 
${\bf S}_j$, located at positions ${\bf R}_j$. 
These local spins   
interact strongly with the itinerant hole spins
via the Kondo--like exchange interaction 
\begin{equation}
\label{Hexch}
H_{\rm{exch}} = \frac{J}{V}
\sum_{j {\bf k} {\bf k'} {\sigma}{{\sigma}^{\prime}}} 
\left(\hspace*{0.05cm} {\bf S}_j \cdot {\bf \sigma} 
\hspace*{0.05cm}\right)_{{\sigma}{\sigma}^{\prime}}
e^{i ({\bf k} - {\bf k'}) \cdot {\bf R}_j} 
\hat{h}^{\dag}_{{\bf k}\sigma} \hat{h}_{{\bf k}^{\prime} {\sigma}^{\prime}},
\end{equation}
where ${\bf \sigma}$ 
is the hole spin operator and $V$ is the volume. 
The coupling of  right--circularly polarized 
optical fields is described in the rotating frame by \cite{sbe}  
\begin{equation}
\label{HL}
H_{L}(t)= -
d(t)  
\sum_{{ \bf k}}
\hat{e}^{\dag}_{{\bf k}\downarrow} \hat{h}^{\dag}_{{-\bf k}\uparrow}
 + h.c, 
\end{equation}
where  $d(t)= \mu_{cv} {\cal E}(t)$ is the Rabi energy, 
$\mu_{cv}$ is the dipole transition matrix element, 
and 
${\cal E}(t) \propto
\exp{[ -t^{2}/\tau_p^{2}]}$
is the optical pulse, with central frequency 
$\omega_p$ and duration $\tau_p$. 
In the above Hamiltonian,  the 
spin quantization z--axis was taken
parallel to the direction 
of optical pulse propagation, which similar to Ref. \cite{oiwa} 
was chosen 
here {\em perpendicular} 
to the  ground state spin polarization. 

The ground state 
of the above Hamiltonian 
is typically described by treating 
the interaction Eq.(\ref{Hexch}) 
within the 
mean field  virtual crystal approximation, which 
neglects spatial correlations 
and assumes 
uniformly distributed classical Mn spins
\cite{mean,review}.  
Within this  approximation, the 
holes experience an effective magnetic field 
proportional to the 
average Mn spin \cite{mean,review}.
The valence states then split into two spin--polarized 
bands,  separated 
by the magnetic exchange energy
$ JcS$, where $c$ is the Mn concentration, and populated by the hole
 Fermi sea. The spins of the above  bands 
point antiparallel and parallel to the Mn spin, 
which is taken here to define the 
x--axis.

We calculated the ultrafast 
nonlinear optical response 
by solving the  equations 
of motion for the density matrix.
For right--circularly polarized light,   spin--$\uparrow$ 
electron states are not populated
for the above Hamiltonian,  
and we only need to consider 
the interband optical polarizations 
$P_{{\bf k} \sigma} 
= \langle \hat{h}_{-{\bf k} \sigma}
\hat{e}_{{\bf k} \downarrow}  \rangle$,
 the carrier 
populations, the average Mn spin 
${\bf S}$, 
and the hole spin ${\bf s}_{{\bf k}}^h$, 
whose components are $ s_{{\bf k} z}^h = \sum_{\sigma} 
\sigma \langle \hat{h}_{{\bf k} \sigma}^\dag \hat{h}_{{\bf k} \sigma}
\rangle$  and $ s_{ {\bf k} +}^h = 
\langle \hat{h}_{{\bf k} \uparrow}^\dag \hat{h}_{{\bf k} \downarrow}
\rangle$, where 
$s_{ {\bf k}+}^h = s_{{\bf k}x}^h + i  
s_{ {\bf k}y}^h$. 
By factorizing 
all higher density matrices, 
we obtain a closed system of equations
analogous to the semiconductor Bloch equations \cite{sbe}  
generalized to include the spin degrees of freedom. 
Due to the selection rules, Eq. (\ref{HL}),  
right--circularly polarized light excites directly the
spin--$\uparrow$ hole--spin--$\downarrow$ electron 
polarization $P_{\uparrow}$: 
\begin{eqnarray} 
&& \left( i \partial_{t} - \Omega_{{\bf k}} - \frac{J c
S_z}{2}
 \right) P_{{\bf k} \uparrow} - \frac{J 
c S_{-}  }{2} 
P_{{\bf k} \downarrow} 
 = 
\nonumber \\ && 
 - d(t) \left[ 1 - \langle \hat{e}^\dag_{{\bf k} \downarrow} 
\hat{e}_{{\bf k} \downarrow} \rangle 
-  \langle \hat{h}^\dag_{{\bf k} \uparrow} 
\hat{h}_{{\bf k} \uparrow} \rangle \right],   
\label{Pup}
\end{eqnarray} 
where $S_{\pm}(t) = S_{x}(t) \pm i S_{y}(t)$ and 
$\Omega_{{\bf k}} =
\varepsilon^{v}_{\bf{k}} 
+ \varepsilon^{c}_{\bf{k}}  - \omega_p  - i/T_2$,
where $T_2$ is the polarization dephasing time.
The rhs of the above equation describes the Pauli--blocking nonlinearity
(Phase Space Filling) \cite{sbe}, determined by the 
equations of motion 
for the carrier populations, with initial condition the 
 Fermi--Dirac distribution of the  hole Fermi sea. 
Additional nonlinearities come from 
the light--induced 
changes in 
the Mn spin  ${\bf S}(t)$ from its ground state configuration, 
discussed below, which result in  a 
time--dependent effective magnetic field 
$ J c {\bf S}(t)$.
The latter gives  a 
time--dependent e--h pair 
energy $\propto S_z(t)$, which renormalizes the spin--polarized hole bands, 
and a time--dependent coupling $\propto S_{-}(t)$ 
between $P_{\uparrow}$ 
and $P_{\downarrow}$:  
 \begin{eqnarray} 
\left(i \partial_{t}-\Omega_{{\bf k}}+\frac{J 
c S_z }{2}
 \right)P_{{\bf k} \downarrow} - \frac{J c
 S_{+}}{2}P_{{\bf k} \uparrow} 
 = d(t) s_{ {\bf k} +}^h.    
\label{Pdown}
\end{eqnarray} 
$P_{\downarrow}$ is  also generated due to the hole spin coherence, 
photoexcited or ground state, 
described by $s_{ +}^h$ (rhs of Eq.(\ref{Pdown})). 
In turn, the hole and Mn spins 
depend on the optical polarizations 
and precess
due to the exchange interaction
as described by the equations of motion     
 \begin{eqnarray} 
&& \partial_t s^h_{{\bf k}}=Jc{\bf S}\times{\bf s}^h_{{\bf k}}+d(t)
\, {\bf h}_{{\bf k}} \ , \ \partial_{t} {\bf S}=J\sum_{{\bf k}}{\bf s}^h_{{\bf k}}
\times{\bf S}. \label{spin} 
\end{eqnarray}
In the above equation, the photoexcited hole spin 
is determined by the nonlinear optical {\em polarizations} 
(and thus the interactions)  
as described  by the vector 
\begin{equation} 
{\bf h}_{{\bf k}} 
= \rm{Im} \left( P_{{\bf k} \downarrow}, 
-i  P_{{\bf k} \downarrow}, 
 P_{{\bf k} \uparrow} \right). \label{h} 
\end{equation} 
During the femtosecond time scales of interest here, 
we neglected for simplicity the much slower 
precession and relaxation of the Mn spin due to the  effective magnetic field
caused by the magnetic anisotropy and the Gilbert damping
\cite{halperin,Sinova}.
The  nonlinear effects 
were treated  non--perturbatively by solving numerically the above 
system of  coupled equations.

Relaxation 
in ferrmagnetic semiconductors  mostly occurs during time scales 
 longer than the pulse duration.
A notable exception 
is the hole spin relaxation, which occurs 
within 10's of fs
due to the strong 
spin--orbit coupling
in the valence band
and the disorder--induced scattering between the different momentum states 
 \cite{zutic}. 
Carrier spin relaxation is typically described 
within the spin Bloch equations by 
introducing relaxation terms 
so that the spin relaxes
to a direction opposite to the Mn-spin \cite{zutic,halperin}. 
However, this spin relaxation  also leads to strong dephasing of the 
optical polarizations and  coupling between 
$P_{\uparrow}$ and $P_{\downarrow}$. 
Thus, the dephasing of both the  spin and the optical polarizations 
must be treated 
consistently, 
including the light--induced nonlinear 
contributions.
We treated all the above 
effects by using
the Lindblad semigroup description of dissipative quantum dynamics
\cite{lindblad}.
Under the general assumptions of linear coupling between bath and
system operators $L_{{\bf k}}$  
and positivity 
and semigroup--type density matrix time evolution, 
the relaxation contribution 
is given by \cite{lindblad}
\begin{eqnarray}
\label{lind}
&&\partial_t\rho|_{\rm{rel}} =  
 \sum_{{\bf k} {\bf k}'} 
\frac{\Gamma_{{\bf k}{\bf k}'}}{2} \langle 2 
 L_{{\bf k}}  \hat{\rho} L_{{\bf k}'}^{\dag}
- L_{{\bf k}'}^{\dag}
 L_{{\bf k}} \hat{{\rho}} - \hat{\rho} 
L_{{\bf k}'}^{\dag} L_{{\bf k}} \rangle 
\end{eqnarray}
The  Lindblad operators 
$L_{\bf{k}} = \hat{h}^{\dag}_{ {\bf k} \Downarrow} 
\hat{h}_{{\bf k}\Uparrow}$,  
where  $\hat{h}^{\dag}_{\bf{k}\Uparrow}$ 
( $\hat{h}^{\dag}_{\bf{k}\Downarrow}$) 
creates a hole with 
 spin parallel (anti--parallel)
to the Mn spin,
describe relaxation of the  hole spin toward 
the direction opposite to ${\bf S}(t)$ with 
a rate  $\Gamma_{{\bf k}{\bf k}'}
= \delta_{{\bf k}{\bf k}'} \Gamma_{s}$.
The detailed equations, 
obtained 
from Eq.(\ref{lind}) by 
factorizing the higher density matrices, 
will be presented elsewhere.  
Here we note that the hole spin component perpendicular to ${\bf S}$  
dephases with a rate $\Gamma_s$, 
while the spin component 
parallel to ${\bf S}$  
relaxes with a rate $2 \Gamma_s$ 
towards $ 
-s_{\rm{max}}
+ s_{\rm{max}}^2 - ({\bf s}^{h}_{{\bf k}})^2$, 
 where $s_{\rm{max}}$, with 
$s_{\rm{max}}
- s_{\rm{max}}^2= 
 n_{{\bf k}}^h/2   
- (n_{{\bf k}}^h/2)^2$, 
is the maximum hole spin 
for given number of holes $n^h_{{\bf k}}
= \sum_{\sigma} \langle \hat{h}^\dag_{{\bf k} \sigma} 
\hat{h}_{{\bf k} \sigma} \rangle$. 
 Eq.(\ref{lind}) also gives optical polarization 
dephasing and coupling between 
$P_{\uparrow}$ and $P_{\downarrow}$ 
that depend on ${\bf S}(t)$, ${\bf s}^h(t)$, 
and the hole populations,   
whose  expressions 
will be given
elsewhere. We note that electron and Mn spin relaxation can also be 
included by using analogous Lindblad operators.

\begin{figure}
\centerline{
\hbox{\psfig{figure=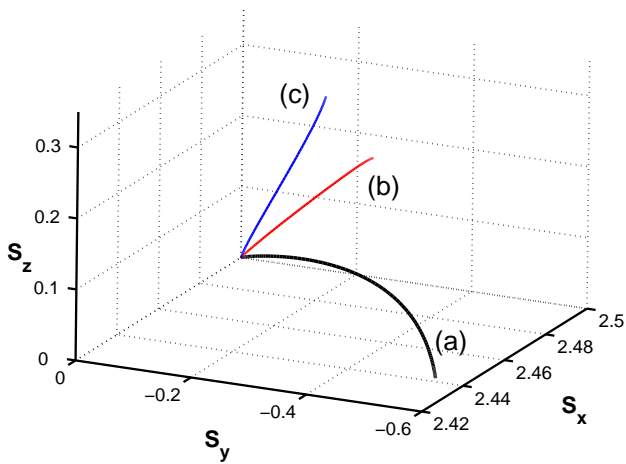,width=8.3cm}}
}
\caption{ Mn-spin trajectory for 
three different relaxation regimes:   
(a) $T_{2}$=2$T_{1}$=330 fs, $\Gamma_{s}=0$,
(b) $T_{2}$=2$T_{1}$=330 fs, $1/\Gamma_{s}$=10.5 fs,
(c) $T_{1}$=165 fs, $T_{2}$=$1/\Gamma_{s}$=10.5 fs. }
\end{figure} 
Fig. 1 shows the light--induced time evolution  of the Mn spin 
for pulse duration $\tau_p = $ 250fs, 
Rabi energy d = 60meV, 
and detuning 10 meV below  the 
threshold of interband absorption. 
Other parameters used in our calculation 
are the exchange energy $JcS$=125 meV,
hole Fermi energy 100 meV, 
fraction of  initial holes 0.33 of the Mn impurities, 
and $m_h/m_e= 7.15$.   
To elucidate the role of  relaxation, 
we compare in Fig. 1 the Mn spin trajectories 
for different values of $\Gamma_s$ and $T_2$. 
The Mn spin
rotates in a {\em clockwise} direction 
within the x--y plane
perpendicular to the direction  of
optical pulse propagation. 
With increasing hole spin damping $\Gamma_s$, 
it develops a large out of plane z--component 
that increases further with decreasing $T_2$. 

The spin dynamics of Fig.1 
comes from the Mn spin
precession around the  mean hole spin. 
Unlike 
for a magnetic field, the latter
changes strongly with time, 
as determined by the
nonlinear optical polarizations, the spin relaxation,
and the precession around ${\bf S}$ 
(Eq.(\ref{spin})). 
For the parameters in Fig.1, 
the hole  spin follows overall the Mn spin trajectory, 
with the exception 
of a component 
perpendicular to ${\bf S}$,  
determined by the nonlinear optical polarizations and the 
spin relaxation, which triggers the Mn spin dynamics.
The light--induced  deviations from the equilibrium 
spin configurations occur {\em within the pulse duration}
and increase with its duration and intensity.

At first glance, the clockwise Mn spin 
rotation is suprising given that, due to the selection rules 
Eq.(\ref{HL}), 
right--circularly polarized light 
creates spin--up holes 
that should lead to {\em counter--clockwise} 
rotation (see Eq.(\ref{spin})). 
In fact, our numerical results show that 
the z--component of the photoexcited 
hole spin is negative due to the exchange interaction. 
This is a consequence of 
the fact that 
${\bf s}^h$ is photoexcited via the nonlinear optical 
{\em polarizations} 
and not by the optical field directly
(see Eq.(\ref{spin})).
Also striking is the strong effect 
of the hole spin relaxation 
on the  Mn spin 
trajectory: it leads to 
the development, {\em within the pulse duration}, 
of a large component $S_z$  in the 
direction of optical pulse propagation despite 
the absence of any Mn spin damping in Eq.(\ref{spin}).
A very fast 
Mn spin tilt toward the z--direction 
was observed experimentally \cite{oiwa}, 
however for weaker photoexcitation. 
The spin dynamics in Fig.1 is significant for Rabi energies
larger than a few tens of meVs, which corresponds to stronger 
photoexcitation than in Ref. \cite{kimel-04}
but not  as strong as in Ref. \cite{wang-05},
 where heating effects \cite{bigot} become pronounced. 

\begin{figure}
\centerline{
\hbox{\psfig{figure=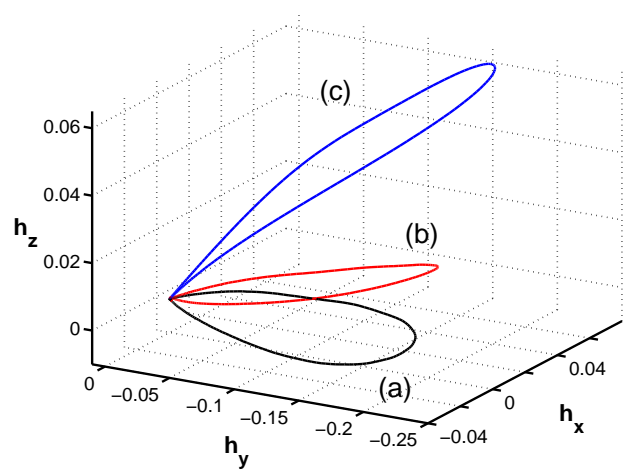,width=8.3cm}}
}
\caption{ Trajectories of the light--induced 
 field $\bf{h}$ (units of $d(0)/2E_{F}$) 
for the three relaxation regimes of Fig. 1.} 
\end{figure}
To interpret  the above results, we 
consider the stationary (adiabatic following) limit, 
where the  hole spin precesses and/or relaxes 
much faster than 
the pulse duration and period 
of Mn spin 
precession,  
and can thus adjust to the instantaneous $d(t)$ and ${\bf S}(t)$ values. 
Neglecting 
the component of 
$\partial_t {\bf s}^h$
perpendicular to ${\bf S}$,  
we can then eliminate the 
hole spin from Eq. (\ref{spin}) 
and obtain after some algebra 
and noting that 
${\bf S} \cdot \partial_t {\bf S}= 0$
 \begin{eqnarray} 
\partial_{t} {\bf S}=
\frac{J d(t)}{(JcS)^2 + \Gamma_s^2} 
 \left[ 
\Gamma_s \ {\bf h}\times{\bf S} - 
J c \
{\bf S}\times\left( {\bf S}\times{\bf h} \right) \right],
\label{S-rel}
\end{eqnarray}
where 
${\bf h}(t) = \sum_{{\bf k}} h_{{\bf k}}$ 
is an effective magnetic field 
determined by the nonlinear optical polarizations 
(Eq.(\ref{h})). 
The above equation for the Mn spin 
demonstrates a light--induced precession around $ {\bf h}(t)$ 
and Gilbert damping--like relaxation of ${\bf S}$ 
toward ${\bf h}(t)$. Such ultrafast nonlinear optical effects  
should be contrasted to the static effects 
governed by the  magnetic anisotropy 
\cite{Sinova,halperin}. 

The light--induced effective  magnetic field ${\bf h}(t)$, Eq.(\ref{h}), 
that governs  the 
Mn spin dynamics has z--component 
determined by 
the dissipative part of the nonlinear optical polarization, 
Im$ P_{\uparrow}$,  
and in--plane components 
determined by $P_{\downarrow}$.  
Its time--dependence is shown in Fig.2. 
The exchange interaction plays an important role  
in determining ${\bf h}(t)$.
To see this, we note that, for $J=0$, 
 $P_{\downarrow}=0$. 
The corresponding ${\bf h}$ then points along the 
z--axis, with magnitude determined by the 
dephasing of $P_{\uparrow}$ and, for $T_2 < \tau_p$, 
proportional to the optical pump rate. 
In this case, the Mn spin would relax to the z--axis.  
However, 
the exchange interaction and spin relaxation 
drastically affect the optical 
polarizations and thus ${\bf h}(t)$.  
For example, 
by coupling the hole spin--$\uparrow$ and spin--$\downarrow$ 
states, 
they 
lead to a large  $P_{\downarrow}$ and thus  
a large x--y plane component of the 
effective magnetic field ${\bf h}$.
This is demonstrated in Fig.2, 
which  shows ${\bf h}(t)$  for different values 
of the relaxation rates. 
In fact, for sufficiently long $T_2$ 
and detunings close to or below the interband absorption threshold,  
Re$P_{\downarrow}$  exceeds 
Im$P_{\uparrow}$, in which case the 
light--induced magnetic field  ${\bf h}$ 
 points mostly along the x--y plane, 
in a direction almost perpendicular to the ground state 
magnetization (x--axis) (see Fig.2(a)).  
For $\Gamma_s=0$, the
second term on the rhs of Eq.(\ref{S-rel}) then leads to 
Mn spin relaxation mostly within the x--y plane, 
while with increasing $\Gamma_s$, the first term leads to 
 precession out of the x--y plane, 
consistent with the numerical results of Fig.1. 
The polarization 
dephasing, due to the hole spin damping as well as the disorder and other 
contributions, enhances the relative magnitude of 
Im${\bf P}_{\uparrow}$ and thus $h_z$ (see Figs.2(b) and 2(c)). 
As a result of $h_z$, 
${\bf S}$  develops an additional z--component as it relaxes 
toward ${\bf h}$. 
One should note that 
the rhs of Eq. (\ref{S-rel})
vanishes after the optical pulse is gone 
and thus the light--induced Mn spin relaxation and precession 
only occurs during the pulse.
Furthermore, the optical polarizations, and therefore ${\bf h}(t)$, 
depend on the spins, and thus 
the Mn spin dynamics
leads to 
an additional time--dependence of ${\bf h}(t)$
and to corresponding nonlinearities. 
This can be seen in Fig.2, where the trajectories of ${\bf h}(t)$ 
have a loop--like shape. Finally, the Phase Space Filling optical nonlinearity
also affects the time--dependence of ${\bf h}(t)$, Eq.(\ref{h}), 
and thus the Mn spin 
relaxation Eq.(\ref{S-rel}). 

To gain  insight into the effects of the optical 
excitation and exchange interaction on the hole spin dynamics,  
one can  solve analytically the polarization
equations of motion, Eqs.(\ref{Pup}) and (\ref{Pdown}),
for $T_2 < \tau_p$
in the stationary limit, 
discussed elsewhere.
For example, 
we obtain this way 
for the x--y spin components  
($\Gamma_s=0$)
\begin{eqnarray} 
\partial_{t} s^h_{{\bf k} +} 
= J \left[ 1 + \frac{d^2(t)/2}{\Omega_{{\bf k}}^2 - (JcS/2)^2} \right]  
{\bf S}\times{\bf s}^h_{{\bf k}}|_{+} \nonumber \\
+ \frac{i d^2(t) \Omega_{{\bf k}}}{
\Omega_{{\bf k}}^2 - (JcS/2)^2}
 \left[ 
s^h_{{\bf k} +} + \frac{J ( 1 - n_{{\bf k}}/2)}{2 \Omega_{{\bf k}}}
S_{+} \right]  \label{s-nonl} 
\end{eqnarray}
where $n_{{\bf k}} 
= \sum_{\sigma} \langle \hat{h}^\dag_{{\bf k} \sigma} 
\hat{h}_{{\bf k} \sigma} \rangle$. 
The first term on the rhs 
describes
hole spin precession around ${\bf S}(t)$
triggered by  a time--dependent effective exchange interaction.  
The second term  describes a 
light--induced in--plane rotation of the hole 
spin, even for $S_z=0$, 
while its dissipative part 
describes a light--induced hole spin relaxation.
Finally, we note that, for very strong photoexcitation \cite{wang-05}
and/or detunings well above the absorption edge, 
the heating of the hole Fermi sea can become significant. 
Similar to  metals \cite{bigot}, 
such hot hole  effects can be treated 
by introducing a time--dependent effective temperature.

In conclusion, we presented a theory of ultrafast 
nonlinear optical and spin dynamics in 
III(Mn)V ferromagnetic semiconductors.
We demonstrated 
a light--induced Mn spin precession and relaxation, 
which can lead to a magnetization tilt of 
several tens of degrees. 
We showed that the latter is
determined by the nonlinear optical polarizations 
and depends critically on the exchange interactions, dephasing, 
and hole spin relaxation. 
The nonlinear light--induced magnetic
effects discussed here occur during the optical pulse and  
can be observed with ultrafast 
pump--probe magneto--optical spectroscopy. 

This work 
was supported by the EU Research Training Network  HYTEC
 (HPRN-CT-2002-00315).


\end{document}